\documentclass[twocolumn,showpacs,preprintnumbers,prb]{revtex4}
\usepackage{graphicx,bm,amsmath,amssymb}
\usepackage{diagbox}
\usepackage{color}

\begin{document}

\title{Impact of capacitance and tunneling asymmetries on Coulomb blockade edges and Kondo peaks 
in non-equilibrium transport through molecular quantum dots}
\author{A. A. Aligia}
\affiliation{Centro At\'{o}mico Bariloche and Instituto Balseiro, Comisi\'{o}n Nacional
de Energ\'{\i}a At\'{o}mica, 8400 Bariloche, Argentina}
\author{P. Roura-Bas}
\affiliation{Dpto de F\'{\i}sica, Centro At\'{o}mico Constituyentes, Comisi\'{o}n
Nacional de Energ\'{\i}a At\'{o}mica, Buenos Aires, Argentina}
\author{Serge Florens}
\affiliation{Institut N\'eel, CNRS et UJF, BP 166, 38042 Grenoble Cedex 9,
France}
\date{\today }

\begin{abstract}
We investigate theorerically the non-equilibrium transport through a molecular quantum 
dot as a function of gate and bias voltage, taking into account the typical situation 
in molecular electronics. In this respect, our study includes asymmetries both in the 
capacitances and tunneling rates to the source and drain electrodes, as well as
an infinitely large charging energy on the molecule. Our calculations are based on the 
out-of-equilibrium Non-Crossing-Approximation (NCA), which is a reliable technique in the 
regime under consideration. We find that Coulomb blockade edges and Kondo peaks
display strong renormalization in their width and intensity as a function
of these asymmetries, and that basic expectations from Coulomb blockade theory
must be taken with care in general, expecially when Kondo physics is at play.
In order to help comparison of theory to experiments, we also propose a simple phenomenological
model which reproduces semi-quantitatively the Coulomb blockade edges that were numerically computed
from the NCA in all regimes of parameters.
\end{abstract}

\maketitle


\section{Introduction}

\label{intro}

In the last decade, an enormous amount of research in the field of
nanoelectronics was devoted to the study of electronic transport through 
quantum dots. A seminal and very versatile type of device is the gate-defined
artificial atom created at the two-dimensional electron gas confined in
semiconducting heterostructures.\cite{gold1,cro,gold2,wiel} More recently, 
molecular quantum 
dots~\cite{liang,parks,parks2,NatelsonReview,serge,urdampilleta,vincent,vanderzant} 
have gained an important
momentum, due to the possibility of playing with the chemistry
of the molecule, which compensates the reduced gate tunability of these systems.
In all these quantum dot setups, several physical effects are generically 
observed when the system is cooled at cryogenic temperatures: Coulomb blockade is
probably the major one, due to the large charging effects in these nanoscopic
quantum dots. In addition, more subtle correlations show up due to the Kondo 
effect~\cite{Hewson,PustilnikGlazman}, which implies a resonance at the Fermi energy in 
the spectral density of the dot state, which in turn leads to an anomalous peak in the 
differential conductance $G(V)=dI/dV$ at zero bias voltage $V$.

However, important and unexpected correlation effects can also take place in
transport features at finite bias voltage. For instance, Coulomb blockade
edges were shown by K\"{o}nemann \textit{et al.}~\cite{haug} to present a strong width
renormalization in the situation of large tunnel asymmetries between source and
drain electrodes, that is also quite typical in molecular quantum
dots.\cite{park} Enhancement of cotunneling 
features of excitation levels (as precursor of the Kondo effect) were also described 
by Paaske \textit{et al.}~\cite{Paaske} and later widely observed. All these features need
to be incorporated in the current theoretical developments of non-equilibrium
quantum transport, which has so far mainly focused in analyzing the main
universal Kondo signatures in $G(V)$ under many simplifying assumptions
(such as symmetric capacitive and tunnel coupling to the electrodes).
In particular, the experiment by K\"{o}nemann \textit{et al.}~\cite{haug}
investigated a strongly confined semiconducting quantum dot coupled to two conducting 
leads with highly asymmetric tunnel couplings, and studied tunneling resonances at 
nearly opposite values of $V$. 
While in absence of interaction effects one expects nearly the same intensity and width of 
the two resonances (see Section~\ref{rele}), the Coulomb blockade edge at positive voltage 
was however nearly five times more intense 
and nearly two times narrower than the peak at negative voltage. 
The authors assumed that for negative voltage, the width in the spectral 
density of the dot level is two times larger than for positive voltage.
This is based on the fact that the
charge-transfer peak (the one near the dot level $E_{d}$) in the density of states 
of the Anderson
model is two times wider when the dot is occupied.\cite{pru,logan}
However, as we shall show, the peak for negative voltage corresponds to the intermediate
valence (IV) regime, in which the width of the spectral density is approximately 3/2 times
larger than in the empty-orbital regime.
In addition, the ratio of widths for negative and positive voltages depends also
on the asymmetries of the source and drain capacitances.
Our calculation will also be able to take into account the presence of a Kondo resonance 
at small bias, which in previous Coulomb blockade theory was clearly lacking.\cite{haug} 

In this paper, after a brief discussion of what is expected in the absence
of Coulomb interactions for asymmetric coupling to the leads, we calculate the
differential conductance $dI/dV$ in the Keldysh formalism within the
non-crossing approximation (NCA).\cite{nca,nca2} The NCA reproduces well the
scaling relations with temperature $T$ and bias voltage in the Kondo regime 
\cite{roura} and was for example also successfully used to interpret
experimental results on a controlled crossover between SU(4) and SU(2) Kondo
states driven by magnetic field in a nanoscale Si transistor \cite{tetta},
as well as quantum phase transitions involving singlet and triplet states
in molecular quantum dots.\cite{serge} 
It was also used recently to discuss non-equilibrium Kondo 
spectroscopy in a system of double quantum dots.\cite{oks} 
For asymmetric coupling to the leads, we will also derive an analytical expression 
which describes semiquantitatively the NCA results. This expression is used to 
interpret the experiments of K\"{o}nemann \textit{et al.}~\cite{haug} providing a good
agreement, and could be used for further studies of molecular systems.

The paper is organized as follows. In Section \ref{model} we present the 
Hamiltonian and the equation describing the current in presence of a finite bias. 
In Section \ref{rele} we compute for comparison the differential conductance 
$G(V)$ in absence of Coulomb interaction.
Section \ref{ncares} contains the NCA results for the non-linear conductance, 
and presents an analysis of the positions and widths of the peaks displayed in 
$G(V)$ for different experimentally relevant conditions.
In Section \ref{compa} we discuss the experiment of K\"{o}nemann \textit{et
al.}, and end up in Section \ref{sum} with a summary.

\section{Model}
\label{model}

We consider the Anderson impurity model with infinite on-site repulsion $U$, an
assumption that is realistic for most molecular quantum dots, but also well obeyed
for very small semiconducting devices in the vicinity of a charge-degeneracy
point. The impurity states thus involve a singlet
configuration $|s\rangle $ together with a spin doublet $|\sigma \rangle $ 
($\sigma=\uparrow $ or $\downarrow $) corresponding to one additional electron (or
hole) in the dot, taking this particle from the left ($L$) or right ($R$)
conducting leads to which the dot is connected.
The Hamiltonian thus reads:
\begin{eqnarray}
H &=&\sum_{\sigma }E_{d}|\sigma \rangle
\langle \sigma |+\sum_{\nu k\sigma }\epsilon _{\nu k}c_{\nu k\sigma
}^{\dagger }c_{\nu k\sigma } \notag \\
&&+\sum_{\nu k\sigma }(V_{k}^{\nu }|\sigma \rangle \langle s|c_{\nu k\sigma
}+\mathrm{H.c}.), \label{ham}
\end{eqnarray}%
where the constraint $|s\rangle \langle s| + \sum_\sigma
|\sigma \rangle \langle \sigma |=1$ is imposed.
Here $c_{\nu k\sigma }^{\dagger }$ create conduction states in the lead $\nu $ 
with spin $\sigma $ and wave vector $k$. The tunnel couplings of the
quantum dot to the leads are $\Delta _{\nu }=\pi \sum_{k}|V_{k}^{\nu }|^{2}\delta
(\omega -\epsilon _{\nu k})$, assumed independent of energy $\omega $, and
we define the total hybridization $\Delta =\Delta _{L}+\Delta _{R}$. We take 
the sign of the bias
voltage in such a way that $\mu _{L}-\mu _{R}=eV$, where $\mu _{\nu }$ is
the chemical potential of lead $\nu $. Note that the interchange of right and
left leads or electrons by holes is equivalent to a change of sign of $V$.
We take the arbitrary origin of energies at $\mu _{R}=0$. The capacitance effects
modify the energy necessary to add an electron to the dot with the lever arm
parameter $\alpha$ (which depends on the source, drain and gate 
capacitances)~\cite{haug,park}, defined as:
\begin{equation}
E_{d}=E_{d}^{0}+\alpha eV, \label{Ed}
\end{equation}%
where $0\leq \alpha \leq 1$, and $e$ is the electron charge (taken positive). For 
the situation of symmetric voltage drop that is usually considered, $\alpha $ is often taken 
near 1/2, but this may not be the case in many experimental setups.

We obtain the conductance $G=dI/dV$ differentiating the current $I$
(numerically from NCA results), which is given by \cite{meir}
\begin{eqnarray}
I &=&C\int d\omega \rho (\omega,V,E_d)[f_{L}(\omega )-f_{R}(\omega )], \notag \\
C &=&\frac{8\pi e\Delta _{L}\Delta _{R}}{h\Delta }, \label{i}
\end{eqnarray}%
where $f_{\nu }(\omega )=f(\omega -\mu _{\nu })$ is the Fermi distribution in
each lead, with $f(\omega )=1/(e^{\omega /kT}+1)$, and $\rho (\omega ,V,E_{d})$
is the non-equilibrium (voltage-dependent) spectral density of the impurity level.

\section{Non-interacting resonant-level model}
\label{rele}

Before discussing in more detail the complex effects of Coulomb interactions, 
we present in this section a simplified model that sheds light onto the expected 
behavior of the differential conductance $G(V)$.
The first assumption that we make in this section is that the tunnel couplings to 
the leads are very asymmetric. This hypothesis is realistic for molecular quantum 
dots and also in the experiment of K\"{o}nemann \textit{et al.} in 
Ref.~\onlinecite{haug}.
We make here the choice $\Delta _{R} \gg \Delta _{L}$, so that 
the system is in equilibrium with the right lead and the explicit dependence on 
the voltage of the spectral density $\rho (\omega )$ disappears [the dependence of
$\rho(\omega)$ with $V$ through $E_d$ given by Eq. (\ref{Ed}) remains]. Then one has
\begin{equation}
\rho (\omega ,V,E_{d})\simeq \rho _{R}(\omega ,E_{d}), \label{rhoap}
\end{equation}
where $\rho _{R}(\omega ,E_{d})$ is the spectral density of the impurity states 
in equilibrium with the right lead, for a dot energy given by Eq. (\ref{Ed}). 
Using equations (\ref{Ed}-\ref{rhoap}), one obtains at $T=0$
\begin{eqnarray}
G(V)&=& C e \rho _{R}(eV ,E_{d}^{0}+\alpha eV) \notag \\
&+& \alpha C e \int_0^{eV} \!\!\!\!\! d\omega \;\frac{\partial \rho _{R}(\omega ,E_{d}^{0}+\alpha eV)}
{\partial E_d}. \label{g0}
\end{eqnarray}
In the limit of asymmetric capacitances $\alpha \rightarrow 0$, only the first 
term survives and $G(V)$ is just a map of the spectral density 
$\rho _{R}(\omega ,E_{d}^0)$, as is usual with experiments with a scanning
tunneling microscope (STM). Therefore, in this limit,
$G(V)$ has a single peak located at $V=E_{d}^0/e$ (the charge-transfer peak),
and no peak for the opposite bias voltage. In addition, a Kondo peak at $V=0$
can show up in presence of Coulomb interactions, if $-E_{d}^0 \gg \Delta$. 

The main assumption that we make in this section is to take a non-interacting 
resonant level spectral density of width $2\Delta$ at half maximum, $\rho _{R}(\omega ,E_{d})\simeq 
\rho_{0}(\omega -E_{d}),$ with
\begin{equation}
\rho _{0}(\omega-E_d)=\frac{\Delta /\pi }{(\omega -E_{d})^{2}+\Delta ^{2}}. \label{rl}
\end{equation}
This is a crude approximation which is certainly not valid in presence of 
interactions (it misses the important Kondo anomaly), but it can describe 
qualitatively the position and width of the Coulomb peaks in $G(V)$ derived 
from the charge-transfer peak in the spectral density of the dot states, which 
can be renormalized from bare values by the interaction, as we discuss in 
Section \ref{phen}.
Assuming, as in Eq. (\ref{rl}), that $\partial \rho _{R}/\partial E_{d}
=-\partial \rho _{R}/\partial \omega $ [but not $\rho _{R}=\rho _{0}$ yet], 
the second term of Eq. (\ref{g0})
can be easily evaluated and one obtains
\begin{equation}
G(V) \simeq Ce[(1-\alpha )\rho _{R}(eV,E_{d})+\alpha \rho _{R}(0,E_{d})],
\label{g1}
\end{equation}
noting that $E_d$ displays an explicit dependance upon the voltage from
Eq.~(\ref{Ed}).
While the assumption made above for the derivatives of the current is not 
exactly true (even in equilibrium) because of the dependence of the weight and width of 
the peaks with $E_d$, this simple expression reflects a general feature of 
the expected Coulomb peaks, in connection to the charge-transfer peak of 
the equilibrium spectral density $\rho _{R}$. Indeed, now $G(V)$ displays two
finite bias peaks, one that is similar to the STM case and is located at 
voltage $eV=E_d$ [namely $eV=E_{d}^{0}/(1-\alpha )$], and a second ``intermediate 
valence'' (IV) peak at an opposite voltage value such that $E_{d}=E_{d}^{0}+\alpha eV=0$. 
In this latter case, the impurity experiences indeed maximal charge fluctuations
because the level is located exactly on-resonance, and thus one can expect
a different width renormalization than for the STM peak, due to the different influence 
of the Coulomb interaction, as we will demonstrate later on.
For positive $E_{d}^{0}$ (corresponding to a vanishing charge in the dot, also
called the empty orbital regime), the STM peak lies at positive bias and the 
IV peak is located at negative bias, and conversely for negative 
$E_{d}^{0}$ (corresponding to a spin-degenerate odd charge in the dot, also
called the local moment regime).
Thus, if $\rho _{R}=\rho _{0}$ is assumed (which is valid only in absence of
Coulomb interaction), the conductance $G(V)$ displays two peaks as a function of 
voltage: i) an ``STM-like'' peak at $V=E_{d}^{0}/[(1-\alpha )e]$ of 
width $2\Delta /[(1-\alpha )e]$ and maximum intensity $(1-\alpha)eC/\Delta$; ii)
an IV peak at $V=-E_{d}^{0}/(\alpha e) $ of width $2\Delta /(\alpha e) $ 
and maximum intensity $\alpha eC/\Delta$. Therefore, in the absence of Coulomb
interactions, the amplitude and width renormalizations are related, and cannot
reproduce the experimental observations made in Ref.~\onlinecite{haug}.
We now turn to a full microscopic calculation of the problem, without recourse
to the many assumptions made above, that incorporates the correct nature of the 
Coulomb blockade edges and Kondo resonances for all parameters in the device.

\section{NCA results for the interacting model}
\label{ncares}

\subsection{The equilibrium spectral density}
\label{dens}

\begin{figure}[t]
\includegraphics[width=8.cm]{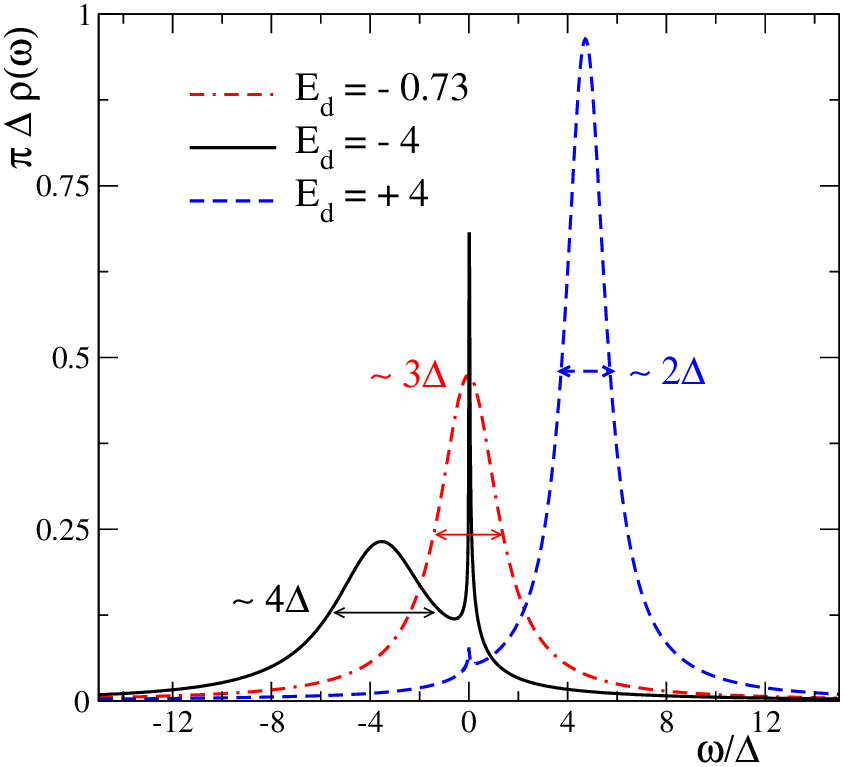}
\caption{(Color online) Equilibrium ($V=0$) spectral density of the 
localized state for the infinite-interaction Anderson model in the empty orbital regime 
(dashed line: $E_d=4 \Delta$, $T=0.01 \Delta$), 
in the intermediate valence (IV) regime (dashed-dot line: $E_d=-0.733 \Delta$, $T= \Delta$),\cite{note} 
and in the local moment regime 
(full line, $E_d=-4 \Delta$,  $T=0.01 \Delta$) that shows an additional Kondo 
feature at low energy.}
\label{rho}
\end{figure}

We consider now Hamiltonian~(\ref{ham}) in its full complexity for the case
of a finite bias voltage. The non-crossing approximation (NCA) 
has proved in the past an excellent technique to deal with large Coulomb
interactions, and can be extended to the present non-equilibrium
situation. In order to gain intuition, we first focus on the zero-bias ($V=0$)
spectral density of the localized level, which is shown on Fig. \ref{rho} for 
low temperatures and three values of the dot energy level $E_d$, corresponding
to the empty orbital ($E_d=4\Delta$), local moment ($E_d=-4\Delta$), and
mixed valence ($E_d \sim 0$) regimes.\cite{note}
For positive $E_d$ the result is very similar to the non-interacting resonant level
prediction, Eq. (\ref{rl}), with a spectral width $2\Delta$, except for the fact 
that the position of the peak is slightly shifted to higher energy 
$E_d^{\text{eff}}=4.72 \Delta$. This energy shift
$E_d^{\text{eff}}-E_d$ is in very good agreement with the Haldane 
shift \cite{hald} $(\Delta/\pi)\text{ln}(D/\Delta)=0.733\Delta$ for the half-band 
width $D=10\Delta$ that we have used in the calculations. This is expected
because the NCA incorporates naturally the lowest order perturbative terms of 
the development in the hybridization $V_k^\nu$.
Note that the small spike at the Fermi energy $\omega=0$ is an artifact of 
the NCA in the empty orbital regime.

In the local moment regime at negative $E_d$, the charge-transfer peak is 
nearly two times wider due to the underlying extra spin degeneracy, as was 
discussed in the introduction and in previous works~\cite{pru,logan}, and 
the NCA correctly reproduces this behavior.
The total intensity is also reduced by correlation effects, as discussed
phenomenologically in Section \ref{phen}, and the position of the peak is also 
shifted to higher energies. More importantly, the screening of the
spin-degenerate states leads to a Kondo peak at the Fermi level, a non-trivial
effect that the NCA is able to describe properly.

Finally, the density of states for the mixed valence situation ($E_d^{\text{eff}}=0$) shows a 
peak centered at the Fermi level as expected, with a width of about $3\Delta$ 
that interpolates between the widths found in the empty orbital and local moment regimes.
The temperature in this calculation has been increased to avoid a shortcoming of the NCA.\cite{note}

Regarding the conductance $G(V)$, these results anticipate similarities and 
differences with the naive expectations based on the non-interacting resonant
level model discussed at the end of the previous section (apart from the
obvious lack of Kondo resonance in the latter model).
A minor point is that $E_{d}^{0}$ should be replaced by an effective level 
$E_{d}^{0,\text{eff}}$, which includes the Haldane shift. 
Based on Fig.~\ref{rho} and Eq.~(\ref{g1}) the width of the STM-like peak in the 
empty orbital regime should be $2\Delta /[(1-\alpha )e]$, thus $4\Delta$ at $\alpha=1/2$ 
for a symmetric voltage drop (as for the non-interacting resonant level model).
However, the width of the STM peak in the local moment regime should rather be 
$4\Delta /[(1-\alpha )e]$, thus $8\Delta$ at $\alpha=1/2$. More subtly,
the IV peak (located at opposite voltage to the STM peak)
should have the same width for the empty orbital and local moment regimes,
namely $3\Delta /[(1-\alpha )e]$, thus $6\Delta$ at $\alpha=1/2$.
These general trends, that we will now examine microscopically from full NCA
calculations of the conductance, show that the experimental analysis of the 
width of Coulomb blockade peak must be done with care.

\subsection{Conductance for asymmetric tunnel coupling to the leads}
\subsubsection{Numerical NCA results}
\label{asym}
\begin{figure}[h!]
\includegraphics[width=8.cm]{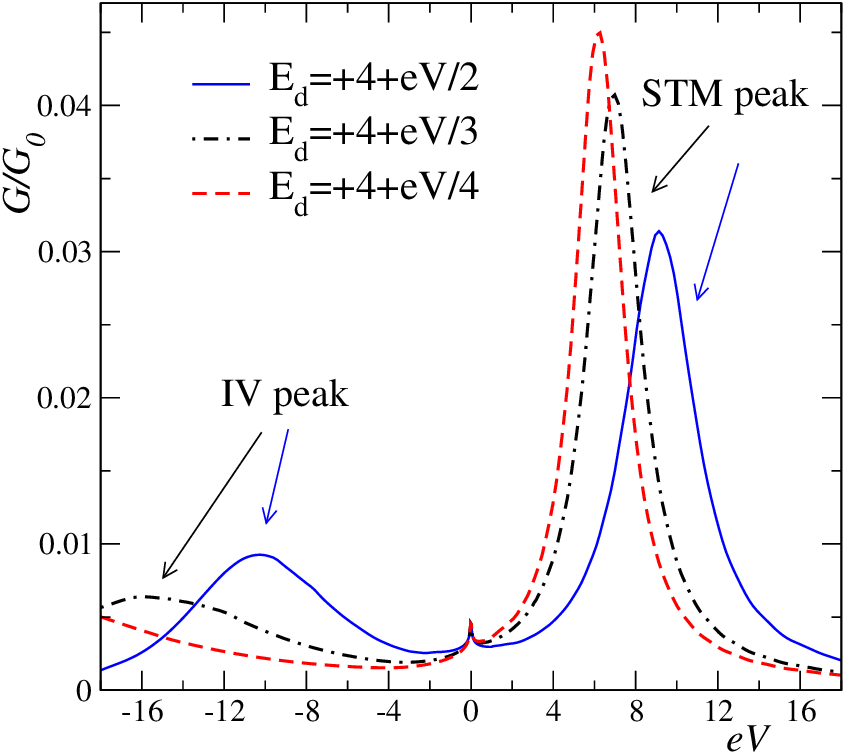}
\caption{(Color online) Finite bias differential conductance $G(V)$ in the empty orbital 
regime $E_{d}^{0}=4 \Delta=4$ for several values of the capacitance ratio 
$\alpha=1/2,1/3,1/4$ defined according to Eq.~(\ref{Ed}), and with an hybridization ratio
$\Delta_R/\Delta_L=60$ (the impurity is here in equilibrium with the right
lead).}
\label{g4}
\end{figure}

In this subsection, we calculate the conductance using the 
non-equilibrium formalism within the non-crossing approximation 
(NCA)~\cite{nca,nca2} for a large choice of parameters. 
Details on the technique are given for example in Ref. \onlinecite{benz}.
We choose here asymmetric tunnel coupling to the leads $\Delta_R=60\Delta_L$, and set the
reference energy scale as $\Delta=\Delta_L+\Delta_R=1$, with a temperature $T=0.01$ 
and a half-bandwidth of the leads $D=10$. Owing to the large ratio
$\Delta_R/\Delta_L\gg1$, the impurity is effectively in equilibrium with
the right lead, although the calculations were in practice performed
using the full out-of-equilibrium NCA.

\begin{figure}[t]
\includegraphics[width=8.cm]{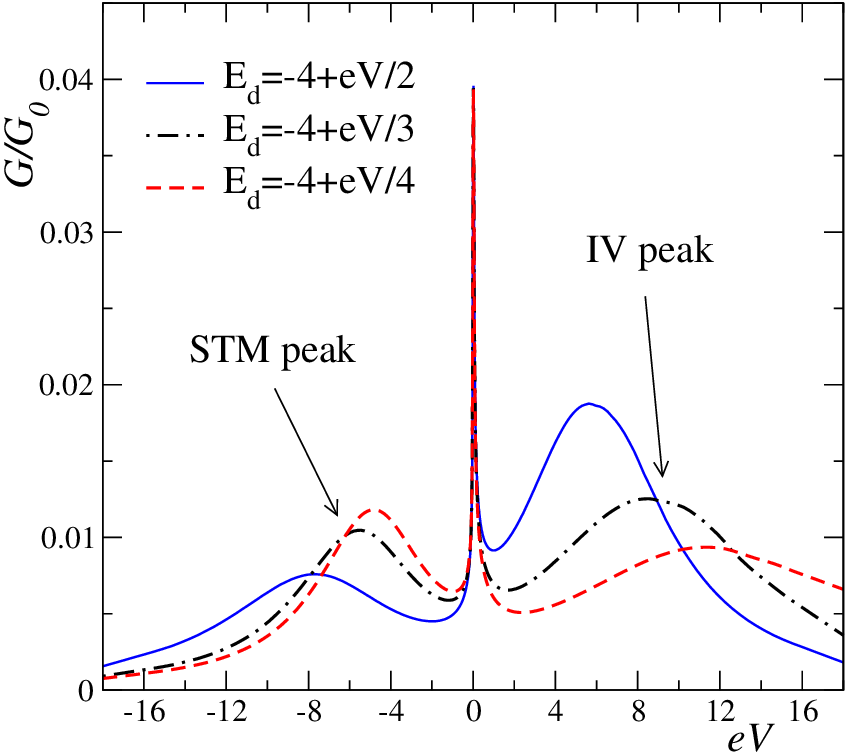}
\caption{(Color online) Finite bias differential conductance $G(V)$ in the local
moment regime $E_{d}^{0}=-4 \Delta=-4$ for several values of the capacitance ratio 
$\alpha=1/2,1/3,1/4$ defined according to Eq.~(\ref{Ed}), and with an hybridization ratio 
$\Delta_R/\Delta_L=60$.}
\label{gm4}
\end{figure}

In Fig.~\ref{g4} we show $G(V)$ in units of the conductance quantum 
$G_0=2e^2/h$ for positive
$E_{d}^{0}$ (empty orbital regime) and several values of capacitance
asymmetries $\alpha$. The small spike at $V=0$ is again an
artifact of the NCA and should be disregarded. For symmetric voltage drop, 
$\alpha=1/2$, the discussion of Section \ref{rele} based on the non-interacting 
resonant level predicts two peaks of equal width $4\Delta$ and equal height 
$eC/(2\Delta)$ at $eV= \pm 2 E_{d}^{0,\text{eff}} \approx 9.4$. However, the more 
refined analysis of Section~\ref{dens}, based on the general features of 
the interacting spectral density, predicts different width renormalization:
a width $4\Delta$ for the STM peak, and a width $6\Delta$ for the
IV peak. These trends are well described by the full NCA
calculation shown in Fig.~\ref{g4}.
However, the conductance $G=dI/dV$ is not directly given by the spectral density, 
but the dependences of its effective width and weight with $E_d$ affect its shape
[second term of Eq. (5)]. This is particularly important for the IV peak, for which
these dependences are larger. This will be disussed in more detail in the next subsection.
The width of the IV peak within the NCA is about 20\% larger than $6\Delta$. This can be due
to the above mentioned effects or to the limitations of the NCA at low temperatures in the
IV regime. In any case, our results demonstrate that the STM and IV peaks are
generically asymmetrical, and that their widths depend sensitively on the charge
state present on the quantum dot.

As $\alpha$ decreases, the tendencies predicted by the simple analysis of Section
\ref{rele} remain qualitatively valid. The position of the STM peak tends to
$E_{d}^{0}$ and its width to $2\Delta$ in the limit $\alpha\to0$. In contrast, 
the IV peak broadens enormously and moves to more negative values 
of $V$ when the capacitances become very asymmetric.

In Fig. \ref{gm4} we show the results for the local moment regime at negative 
$E_{d}^{0}$. In this case, both peaks are much broader than the predictions of 
the naive expectations based on a non-interacting resonant-level spectral density. 
Indeed, the STM peak now lying at negative voltage shows strong interaction effects 
(already witnessed in the spectral density in Fig.~\ref{rho}, that make it twice 
broader than without interactions (the width is $8\Delta$ for symmetric voltage 
drop $\alpha=1/2$).
The IV peak, now located at positive bias voltage, displays
again a width that is about 3/2 wider than the simple expectations outlined at the 
end of Section \ref{rele}, due to combined charge fluctuations and Coulomb
effects (the width is $6\Delta$ for $\alpha=1/2$).
The IV peak is also shifted to the left, since for $\alpha=1/2$ one
would expect it to lie at $eV=2 E_{d}^{0,\text{eff}} \approx 6.5
\Delta$, while it is located at $5.7 \Delta$. 
While naively one would expect a ratio of maxima inversely proportional to the
width, the intensity of the STM peak is smaller than expected, especially for $\alpha=1/2$. 
This further correlation effect is discussed in the next subsection where we
present a more refined analytical approach.
Again, as $\alpha$ decreases, the width and intensity of both peaks evolve
in agreement with the expectations of the non-interacting resonant-level model.

As expected, the most drastic difference with the non-interacting resonant-level model 
for negative $E_{d}^{0}$ is the appearance of the Kondo peak at $V=0$. Note that
its main features are independent of $\alpha$, which will be discussed in Section
\ref{wk}. We now turn to a physical interpretation of the numerical NCA result.

\subsubsection{Phenomenological approach}
\label{phen}

The main reason for the failure of the non-interacting resonant-level model to account 
for Coulomb blockade peaks at asymmetric tunnel coupling to the leads is
that the correct width and weight of the peak in the impurity level spectral density 
$\rho(\omega)$ both depend in a non-trivial way on the occupancy, as discussed in 
Section \ref{dens}.
One of the simplest ways to modelize this effect is the alloy analogy approach to periodic 
Anderson models, in which the electrons with spin down are frozen to calculate the dynamics 
of electrons with spin up and vice versa.\cite{aaa,dyn} This approach is also
related to the so-called Hubbard-III approximation,\cite{hub3} in which the weight of the 
charge-transfer peak near $E_d$ for spin $\sigma$ (or the lower Hubbard band in a 
Hubbard model) is proportional to the number of unoccupied states with the opposite 
spin, $1-\langle n_{-\sigma }\rangle$.

To correct the width inconsistency in the non-interacting resonant level model, we have
fitted the width of the NCA peak as a function of $E_d$ at a temperature 
$T=2 \Delta$ (such that the Kondo peak is absent and the width of the IV 
peak is well behaved \cite{note}), with other parameters taken as 
in Section \ref{asym}. 
For simplicity, in this section and in section \ref{compa}, we replace the
notation $E_d^{\text{eff}}$ by $E_d$, incorporating the Haldane shift in the 
latter. From this, we obtain that the effective half-width at half-maximum 
can be approximated as:
\begin{eqnarray}
\frac{\Delta _{\text{eff}}}{\Delta } &=&\frac{3}{2}-\frac{1}{2}\tanh \left( 
\frac{E_d}{2.77\Delta}\right). 
\label{deff}
\end{eqnarray}
This simple formula nicely interpolates between the empty orbital case
($\Delta _{\text{eff}}=\Delta$ for $E_d\gg\Delta$) and the local
moment regime ($\Delta _{\text{eff}}=2\Delta$ for $-E_d\gg-\Delta$),
passing through the mixed valence regime ($\Delta _{\text{eff}}
=(3/2)\Delta$ for $|E_d|\ll\Delta$).

The above discussion suggests that a better description of the single peak 
in the equilibrium local density of state can be obtained using the improved
formula:
\begin{equation}
\rho _{R}(\omega ,E_{d})=\frac{(1-\langle n_{\sigma }
\rangle )\Delta _{\text{eff}}/\pi }{(\omega -E_{d})^{2}+\Delta _{\text{eff}}^{2}}, \label{rf}
\end{equation}%
where $\langle n_{\sigma }\rangle =\int d\omega \rho _{R}(\omega)f_{R}(\omega )$
is the occupation of the dot for a given spin, independent of the spin in the
absence of an applied magnetic field. Integration gives $\langle
n_{\sigma }\rangle $ in terms of the digamma function $\psi (x)$:
\begin{eqnarray}
\langle n_{\sigma }\rangle &=&\frac{1/2-\Psi _{R}}{3/2-\Psi _{R}}, \notag
\\
\Psi _{\nu } &=&\frac{1}{\pi }\text{Im}\psi \left[ \frac{1}{2}+\frac{\Delta
_{\text{eff}}+i(E_{d}-\mu _{\nu })}{2\pi T}\right] . \label{ns}
\end{eqnarray}
Replacing then Eq. (\ref{rf}) in Eq. (\ref{i}) gives the current:
\begin{equation}
I=C(1-\langle n_{\sigma }\rangle )(\Psi _{R}-\Psi _{L}), \label{if}
\end{equation}
Finally differentiating this expression, one obtains an analytical although 
lengthy expression for the differential conductance $G(V)$, which we do not 
reproduce here.
At $T=0$, $G(V)$ contains two terms which have the form of those included in 
Eq. (\ref{g1}), plus some additional terms due to the dependence of 
$\Delta _{\text{eff}}$ and $\langle n_{d\sigma}\rangle $ with respect 
to $E_{d}$. These additional terms affect the IV peak, 
for which the variation of $\Delta _{\text{eff}}$ and $\langle n_{d\sigma }\rangle $ 
is significant.

\begin{figure}[t]
\includegraphics[width=8.cm]{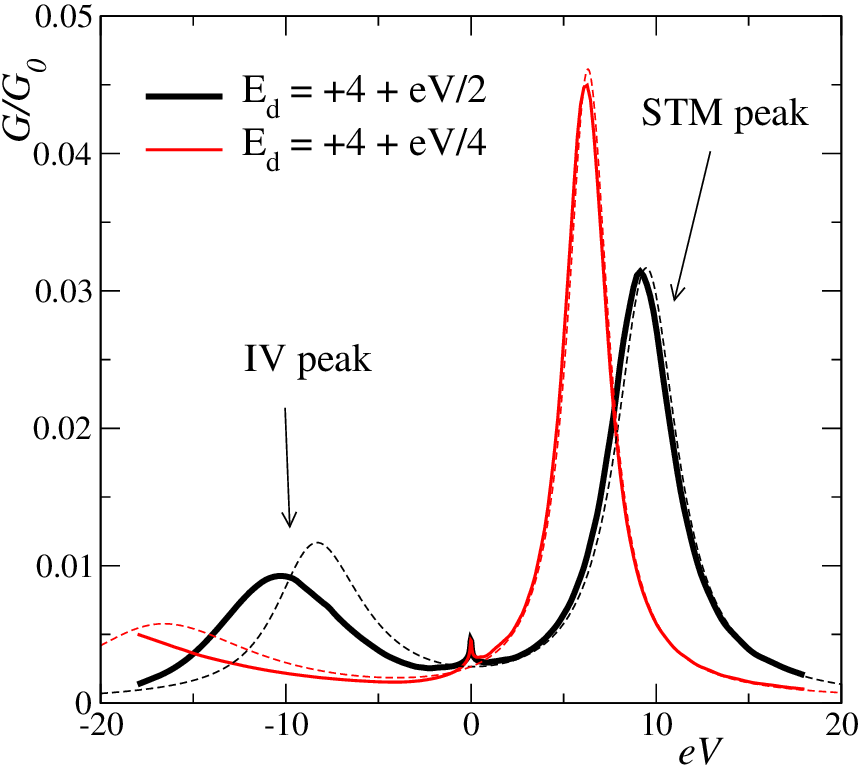}
\caption{(Color online) Differential conductance in the empty orbital regime 
$E_{d}^{0}=4 \Delta=4$, for asymmetric tunnel couplings ($\Delta_R = 60 
\Delta_L$), low temperature $T=0.01 \Delta$, and two values of the capacitance 
ratio $\alpha=1/2,1/4$. Full lines are the NCA simulations, and dashed lines are
the results from the phenomenological approach.}
\label{compa+}
\end{figure}

In Fig. \ref{compa+} we compare the conductance $G(V)$ obtained through this
phenomenological approach to the full NCA result in the empty orbital case 
$E_{d}^{0}=4 \Delta$ (note that the Haldane shift is included by hand in the
level position within the phenomenological approach).
We remind the reader that for $\alpha=1/2$, the non-interacting resonant-level model predicts a 
symmetric $G(V)=G(-V)$ conductance curve, which turns out to be incorrect
in the presence of strong Coulomb interaction.
The phenomenological approach is thus a considerable improvement for the
IV peak at negative voltage, although some quantitative
discrepancies remain. In particular, the analytical expression gives an
intermediate valence peak a little bit narrower and shifted towards $V=0$,
but the general trends are well reproduced by the simple approach.
In addition, this discrepancy might be due to limitations of the NCA
at low temperatures in this regime.

\begin{figure}[t]
\includegraphics[width=8.cm]{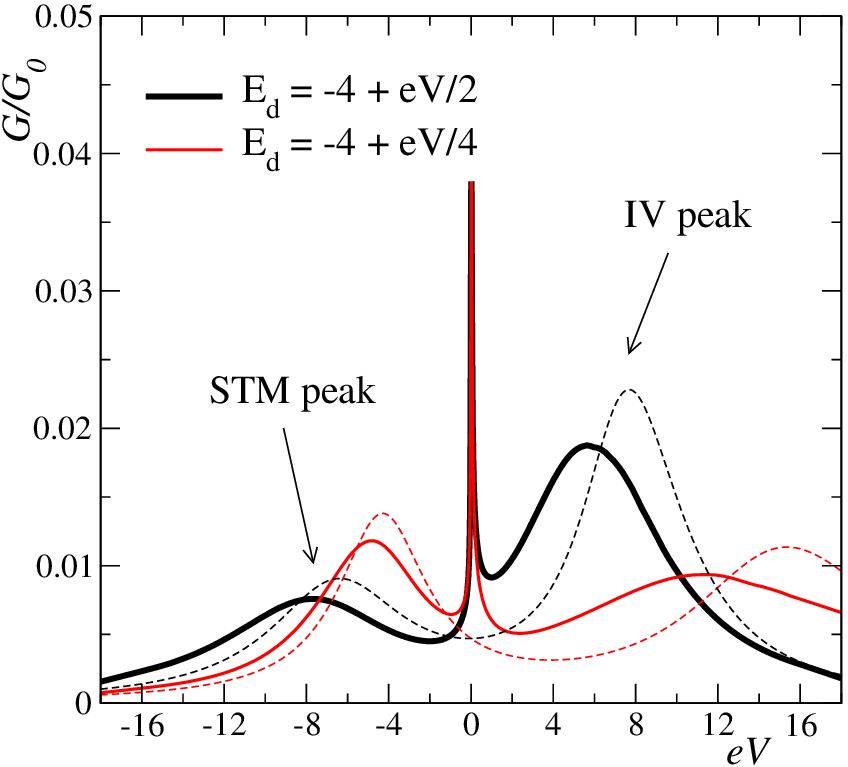}
\caption{(Color online) Differential conductance in the local moment regime 
$E_{d}^{0}=-4 \Delta=-4$, with the same parameters as in Fig. \ref{compa+}.
Full lines are the NCA simulations, and dashed lines are
the results from the phenomenological approach.}
\label{compa-}
\end{figure}

In the local moment regime for negative $E_{d}^{0}$, as shown in Fig. \ref{compa-}, 
the comparison is less satisfactory. The Kondo peak is obviously missing in the 
phenomenological approach and the positions of the Coulomb peaks show also some 
deviations, particularly for the IV peak now located at positive 
$V$. Nevertheless, the intensities and the widths (which are larger than those 
predicted by the non-interacting resonant-level model) are semiquantitatively reproduced. The small 
discrepancy in the position of the STM peak at negative $V$ might be due to 
some dependence with $E_d$ of the Haldane shift, being smaller for negative $E_d$.
Having clarified the physics at play in the asymmetric situation, which
simplifies because the system is in equilibrium with a single lead, we now
consider the problem in is full complexity for the regime of 
symmetric tunnel couplings.

\subsection{Conductance for symmetric tunnel coupling to the leads}
\label{symme}

\subsubsection{Numerical NCA results}

\begin{figure}[t]
\includegraphics[width=8.cm]{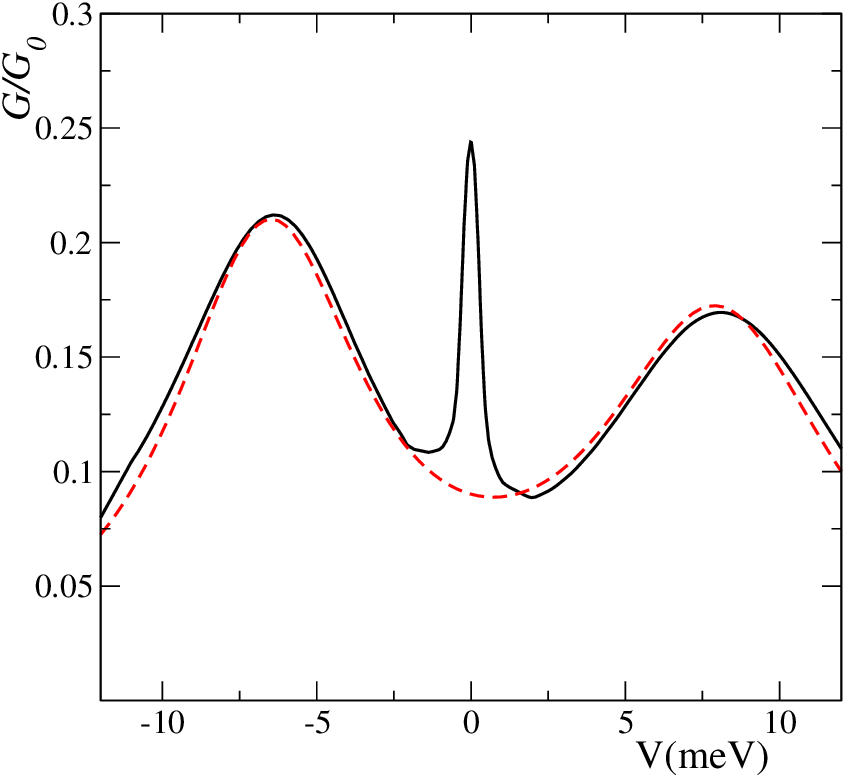}
\caption{(Color online) Differential conductance at symmetric tunnel couplings,
$\Delta_L=\Delta_R$, for parameters close to the experimental measurement
of K\"{o}nemann \textit{et al.}~\cite{haug}, namely
$\Delta=1.15$ meV, $E_{d}^{0}=-4.5$ meV, $T=0.014$ meV and $\alpha=0.425$. 
The full line is the non-equilibrium NCA calculation, while the dashed line
corresponds to a fit to the non-interacting
resonant-level model with renormalized parameters $\Delta^\mathrm{eff}=2.1$ meV, 
$E_{d}^{0}=-3.44$ meV, $\alpha=0.45$ and a scale reduction factor 0.355.}
\label{sym}
\end{figure}

For symmetric tunnel coupling to the leads, an approximation like
Eq.~(\ref{rhoap}), which would permit to reduce the problem to the 
calculation of the spectral density at equilibrium, is no more possible. 
In Fig. \ref{sym} we show the results of the non-equilibrium NCA calculation 
for symmetric tunnel barriers, $\Delta_L=\Delta_R=\Delta/2$,
and parameters similar to those of the experiment by 
K\"{o}nemann \textit{et al.},\cite{haug}
but with negative $E_d^0$.
In contrast to the results of Section \ref{asym}, the symmetry now imposes that 
$G(V)=G(-V)$ for $\Delta_L=\Delta_R$ and $\alpha=1/2$. 

While one would expect that the non-interacting resonant-level model is meaningless in this
non-equilibrium interacting case, the symmetric coupling to the leads ``blurs''
in the conductance the asymmetry due to correlation effects. The simplest
non-interacting resonant level model with renormalized parameters is indeed 
able to reproduce the Coulomb blockade peaks obtained
with the NCA (of course the Kondo anomaly is not captured). 
Except for the small modification of the $\alpha$ value, which is hard to justify, the
changes in the other parameters of the model are expected: a width 4.2 meV nearly two
times larger than expected (2.30 meV) because of correlation effects, an effective $E_d$
shifted upwards due to renormalization effects (the Haldane shift \cite{hald} is
1.04 meV for the half band width $D=20$ meV used in the figure) and smaller
intensity also due to correlation effects (weight proportional to 
$1-\langle n_{-\sigma }\rangle$). 
We now turn to a detailed description of the Kondo resonance for a large 
range of parameters.

\subsubsection{On the width of the non-equilibrium Kondo peak}
\label{wk}

In this section, we study the relation between the width of the low bias
Kondo resonance in the differential conductance $G(V)$ (which is present for negative enough $E_d^0$) 
and the Kondo peak in the equilibrium spectral density of the impurity level $\rho (\omega )$. 
We define $T_K^{G}$ (resp. $T_K^{\rho}$) as the energy scale corresponding to half 
the width at half maximum of the peak in $G(V)$ (resp. $\rho (\omega )$) near $V=0$ 
(resp. $\omega=0$). Both quantities are of the order of the Kondo temperature,
but differ from a non-universal prefactor. Since the Kondo temperature is often
extracted experimentally from the differential conductance curve, but theoretically the
equilibrium density of states is more standardly computed, for instance with the
Numerical Renormalization Group (which has difficulties in grasping the effect
of finite bias voltage), we consider here systematically the ratio 
$T_K^{G} / T_K^{\rho}$, for several ratios of couplings to the leads 
$\Delta _{R} / \Delta _{L}$, keeping $\Delta=\Delta _{R} + \Delta _{L}=1$ fixed, 
and also for several values of the level position $E_d$.
In this section we keep $E_d$ fixed and assume a symmetric voltage drop
$\mu_L=-\mu_R=eV/2$, as $V$ is increased, and take the half band width $D=10$. 
Both the differential conductance $G(V)$ and equilibrium density of states
$\rho (\omega )$ are computed with the NCA, the former with the non-equilibrium 
formalism, and the latter at equilibrium.\cite{nca,nca2,benz}
The resulting ratio of Kondo temperatures is shown in Table \ref{Table1}.

\begin{table}[tbp]
\centering
\caption{Ratio $T_K^{G} / T_K^{\rho}$ of the Kondo temperatures extracted from
finite bias conductance $G(V)$ and equilibrium density of states $\rho(\omega)$,
obtained for different values of the level energy $E_d$ and various
asymmetry ratios $\Delta _{R} / \Delta _{L}$.}
\label{Table1}
\vspace*{0.5cm}
\begin{tabular}{|c|c|c|c|c|c|}
\hline
\diagbox[width=6em,height=3em]{$\Delta _{R} / \Delta _{L}$}{$E_d$} & -2 & $-3$ & $-4$ & $-5$ & $-6$ \\
\hline
1 & 1.59     & 1.45       & 1.33     & 1.31     & 1.30 \\
\hline
2 & 1.50     & 1.19     & 1.078      & 1.09     & 1.08 \\
\hline
5 & 1.29     & 1.03     & 1.008      & 1.003    & $\sim$ 1 \\
\hline
10& 1.29     & $\sim$ 1 & $\sim$ 1   & $\sim$ 1 & $\sim$ 1\\
\hline
20& 1.29     & $\sim$ 1 & $\sim$ 1   & $\sim$ 1 & $\sim$ 1\\
\hline
30& 1.29     & $\sim$ 1 & $\sim$ 1   & $\sim$ 1 & $\sim$ 1\\
\hline
\end{tabular}
\end{table}

\begin{figure}[t]
\includegraphics[width=8.cm]{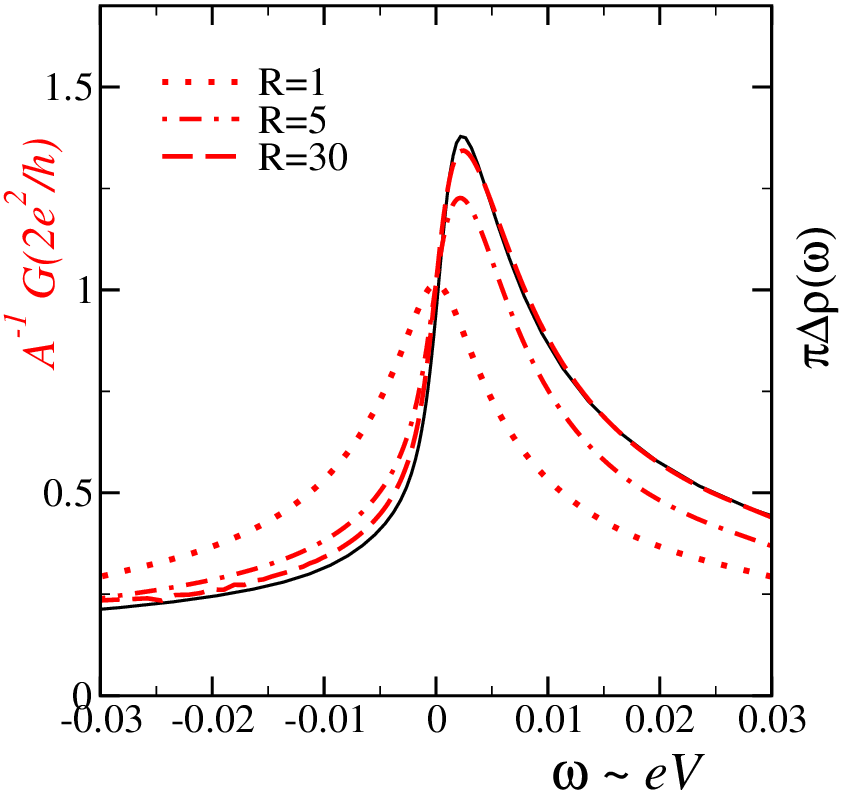}
\caption{(Color online) Differential conductance near $G(V)$
(broken lines in red, corresponding to the left vertical scale), and
equilibrim spectral density of the quantum dot $\rho(\omega)$ (full line in black,
corresponding to the right vertical scale), computed for $\Delta=1$,
$E_d=-4$ and several tunnel asymmetry ratios $R=\Delta_R/\Delta_L=1,5,30$. 
The coefficient $A=4 \Delta_R \Delta_L/\Delta^2$ is introduced to normalize
the conductance to the density of states.}
\label{gdasym}
\end{figure}

It can be noticed from Table \ref{Table1} that the ratio $T_K^{G} / T_K^{\rho} \geq 1$ 
and decreases towards unity when both the tunnel asymmetry ratio $R=\Delta _{R} /
\Delta _{L}$ is increased and the impurity level position $E_d$ is decreased
negatively, which can be understood as follows. 
For large asymmetry of the tunnel coupling to the leads, $R\gg1$, the dot can be 
considered at equilibrium with the right lead, and Eq. (\ref{rhoap}) becomes valid.
Furthermore, in the Kondo limit for fixed chemical potential $\mu$, the Kondo peak is insensitive 
to a small change in the localized level $E_d$, and remains close to the Fermi level. 
For $|eV| \ll T_K^{\rho}$ this is supported by Fermi-liquid 
approaches.\cite{sela,scali} Thus, one can drop the second term in Eq. (\ref{g0})
and $G(V)$ for $|eV|<T_K^{\rho}$ gives directly the spectral density of the dot state $\rho _{R}(eV ,E_{d}^0)$. 
This effect is evident in Fig. \ref{gdasym}, where the evolution of the Kondo
peak in conductance $G(V)$ with increasing asymmetry ratio $R$ is shown and compared 
to the equilibrium density of states $\rho _{R}(eV ,E_{d}^0)$.

\section{Comparison with experiment}
\label{compa}

In this last section, we discuss the experiment of K\"{o}nemann \textit{et al.}.\cite{haug}
This experiment corresponds to a value of $E_{d}^{0}$ that is positive and very large (two orders of 
magnitude bigger than $\Delta$). This situation is very hard for the numerical
solution of the self-consistent NCA equations, but since the Kondo effect does
not take place, we can compare the experiment to the phenomenological approach
developed in Section \ref{phen}, which takes into account the changes in 
both width and height of the charge-transfer peak in the spectral density 
with the dot level $E_d$. The experiment was done for highly asymmetric tunnel coupling 
to the leads,  so that the main assumption of Section \ref{phen} is fulfilled.

\begin{figure}[t]
\includegraphics[width=8.cm]{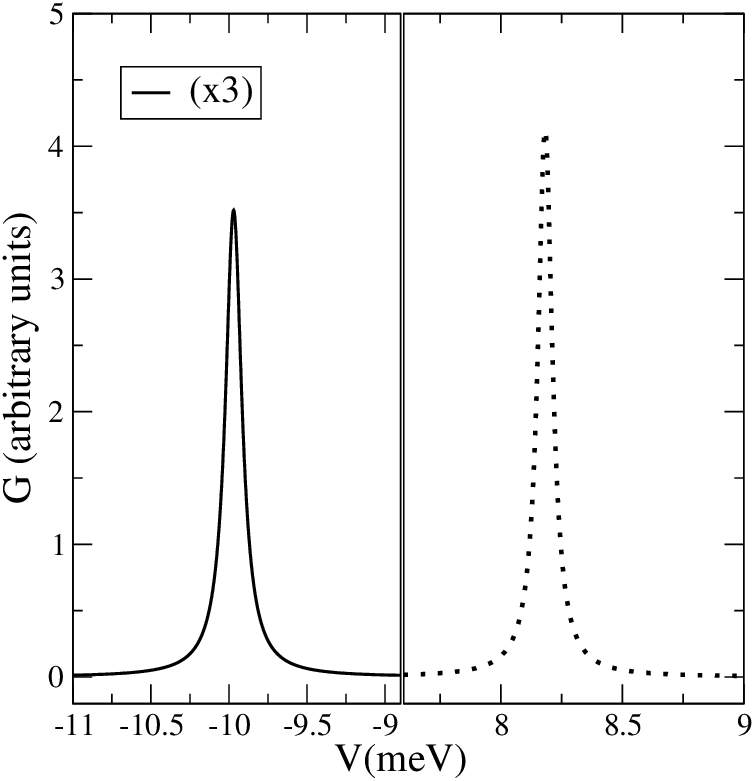}
\caption{(Color online) Differential conductance using the phenomenological
model Eq. (\ref{if}) for parameters corresponding to the experiment by 
K\"{o}nemann \textit{et al.}:\cite{haug}
$\Delta=0.02$ meV, $E_{d}^{0}=4.5$ meV, $T=0.01$ meV and $\alpha=0.45$. Full (dotted) 
line corresponds to negative (positive) bias voltage.}
\label{hau}
\end{figure}

In the experiment, there is a peak at $V=$-9.7 mV with a half width of
76 $\mu$V and another one at $V=$8.9 mV with a half width of 35.5 $\mu$V which 
is about five times more intense.
For $\alpha \simeq 1/2$ (as suggested in Ref.~\onlinecite{haug}), the non-interacting 
resonant level model incorrectly predicts two peaks of equal width and height at 
opposite voltages (see the end of Section \ref{rele}).
For $\alpha=1/2$, our more accurate phenomelogical theory predicts a ratio of about 
3/2 between the width of the IV peak at negative $V$ and the STM peak 
at positive $V$, and not a ratio of about two. However, the fact that the
positions of the two peaks are not symmetric with respect to zero bias
shows that $\alpha$ is in fact different from 1/2. Using $\alpha=0.45$ to adjust
the correct peak position, this provides an additional contribution to the ratio
in the peak widths. 
In Fig. \ref{hau} we display our results for parameters close to the experiment. 
While we are unable to reproduce exactly the intensities, positions and widths of both
peaks, we obtain a semiquantitative agreement with experiment. Our conductance
curves can be fitted with the sum of two Lorentzians: one located at $V=-9.97$
mV with half-width at half-maximum 64 $\mu$V and peak value $1.17A$, another one located
at $V=8.18$ mV, with half-width at half-maximum 36 $\mu$V and peak value $4.21A$.
In the future, comparisons to further experimental data for molecular quantum dots deep
in the Kondo regime would be interesting, in order to test our predictions.

\section{Summary}
\label{sum}
Motivated by experiments with semiconducting and molecular quantum dots~\cite{haug,park}, we have analyzed 
the complete shape of the differential conductance $G(V)$ in the infinite-$U$ Anderson model 
connected to two leads, as a function of the ratio of the potential drops between the dot and 
both leads, and for various tunneling asymmetries. 
The simplest situation is the one found in scanning tunneling spectroscopies,
where both the potential drop and tunneling rates are strongly asymmetric
towards the metallic surface on which a molecule is deposited. As is well-known
in this situation, the conductance provides exactly a measure of the equilibrium spectral 
density of the quantum dot and displays only one Coulomb peak when voltage is aligned to the 
charge-transfer peak in the spectral density (this peak position may contain
renormalization effects). When the impurity level is deeply negative, a local
moment forms, and the Kondo effect leads to an additional anomaly near zero
bias, both in the density of states and differential conductance.

The situation of an arbitrary ratio for the capacitance to the
source and drain is more complex (even still assuming very asymmetric tunneling
rates). We find that the Coulomb peak position and width in this regime not only
display trivial changes expected from the modification of the
capacitance ratio, but are also corrected by many-body effects.
As it is known from previous calculations,\cite{pru,logan}, and reproduced by 
our NCA results, the width of the charge-transfer peak is about two times larger 
for large and negative level position (local moment regime) than for large and 
positive position (empty orbital regime).
In addition, an extra Coulomb peak appears at opposite voltage values to the 
peak found in the STM situation, as expected from Coulomb blockade theory.
However, this peak has in general a different nature, and relates to the
intermediate valence situation of the underlying quantum impurity model,
and displays accordingly a width renormalization of about 3/2 compared to
expectations based on a non-interacting effective resonant level model.
All these effects can be taken into account by a simple phenomenological approach 
described in Section \ref{phen}, which is able to describe well our numerical
simulations (except for the presence of the Kondo resonance) and agrees semiquantitatively 
with some published experimental results [see Section \ref{compa}].

We also provided some general study of the Kondo temperature extracted from
the differential conductance (as typically done in experiments), and made 
some precise connection to the Kondo scale extracted theoretically from
the equilibrium spectral density. We find that the former scale can be up to
60\% larger than the latter one, and that they coincide only for asymmetric 
tunnel couplings deeply in the Kondo regime.

We hope that all these results will serve as guidelines for interpreting 
more quantitatively experimental data obtained from molecular quantum dots,
since many-body effects can quantitatively alter the predictions of basic
Coulomb blockade theory.

\section*{Acknowledgments}

We are partially supported by CONICET, Argentina. This work was 
sponsored by PICT 2010-1060 and 2013-1045 of the ANPCyT-Argentina, 
PIP 112-201101-00832 of CONICET, and PICS CNRS-CONICET 5755.


\begin{thebibliography}{99}
\bibitem{note} In the intermediate valence regime and for small temperatures, the width
of the charge-transfer peak in the spectral density becomes about two times narrower
than expected in the NCA calculations. Therefore we have taken 
$T \ge \Delta$ to calculate correctly its width in this regime.
\end{thebibliography}
\end{document}